\documentclass[preprint2]{aastex}

\newcommand{\eg}{{e.g.,}}
\newcommand{\ie}{{i.e.,}}

\shorttitle{Spectacular Shells in MC2\,1635+119}
\shortauthors{Canalizo et al.}

\begin{document}

\title{Spectacular Shells in the Host Galaxy of the QSO
MC2\,1635+119\altaffilmark{1}}

\author{Gabriela Canalizo\altaffilmark{2,3}, Nicola Bennert\altaffilmark{3},
Bruno Jungwiert\altaffilmark{3,4},
Alan Stockton\altaffilmark{5}, Fran\c{c}ois Schweizer\altaffilmark{6},
Mark Lacy\altaffilmark{7}, Chien Peng\altaffilmark{8}}

\altaffiltext{1}{Based on observations made with the NASA/ESA Hubble Space 
Telescope, obtained at the Space Telescope Science Institute, which is 
operated by the Association of Universities for Research in Astronomy, Inc., 
under NASA contract NAS 5-26555. These observations are associated with 
program \# GO-10421.}

\altaffiltext{2}{Department of Physics and Astronomy, 
University of California, Riverside, CA 92521, USA;
email: gabriela.canalizo@ucr.edu}

\altaffiltext{3}{Institute of Geophysics and
Planetary Physics, University of California, Riverside, CA 92521, USA;
email: nicola.bennert@ucr.edu, bruno.jungwiert@ucr.edu}

\altaffiltext{4}{Astronomical Institute, Academy of Sciences of the Czech 
Republic, Bo{\v c}n\'\i\ II 1401, 141 31 Prague 4, Czech Republic}

\altaffiltext{5}{Institute for Astronomy, University of Hawaii, 2680 
Woodlawn Drive, Honolulu, HI 96822, USA; email: stockton@ifa.hawaii.edu}

\altaffiltext{6}{Carnegie Observatories,
813 Santa Barbara Street, Pasadena, CA 91101, USA; email: schweizer@ociw.edu}

\altaffiltext{7}{Spitzer Science Center, 
California Institute of Technology, Pasadena, CA 91125, USA;
email: mlacy@ipac.caltech.edu}

\altaffiltext{8}{Space Telescope Science Institute, 3700 San Martin Drive,
Baltimore, MD 21218, USA; email: cyp@stsci.edu}

\begin{abstract}
We present deep $HST$/ACS images and Keck spectroscopy of
MC2\,1635+119, a QSO hosted by a galaxy previously classified as
an undisturbed elliptical. Our new images reveal
dramatic shell structure indicative of a merger event in the relatively
recent past.
The brightest shells in the central regions of the host are distributed 
alternately in radius, with at least two distinct shells on one side of the 
nucleus and three on the other, out to a distance of $\sim$13 kpc. 
The light within the five shells comprises $\sim$6\% of the total galaxy 
light.  
Lower surface brightness ripples or tails and other debris extend out to a 
distance of $\sim$65 kpc.  
A simple N-body model for a merger reproduces the inner shell 
structure and gives an estimate for the age of the merger between 
$\sim$30 Myr and $\sim$1.7 Gyr, depending on a range of reasonable assumptions.
While the inner shell structure is suggestive of a minor merger,
the total light contribution from the shells and extended
structures are more indicative of a major merger.
The spectrum of the host galaxy is dominated by a population of 
intermediate age ($\sim$1.4 Gyr), indicating a strong starburst episode 
that may have occurred at the time of the merger event.  We speculate that 
the current QSO
activity may have been triggered in the recent past by either a minor merger, 
or by debris from an older ($\sim$Gyr) major merger that is currently 
``raining'' back into the central regions of the merger remnant.
\end{abstract}

\keywords{galaxies: active -- galaxies: interactions --- galaxies: evolution --- quasars: general --- quasars: individual (MC2\,1635+119)}

\section{Introduction}

The nature of QSO host galaxies has been debated for over four decades.  
Although the terms of the debate have gradually evolved during this time, 
there has been some progress.  We now know, for example, that the 
majority of luminous low-redshift QSOs, whether radio loud or radio quiet, 
reside in the centers of galaxies that have relaxed light distributions like 
ellipticals \citep[\eg][]{dis95,bah97,dun03,flo04}.  This result ties in 
nicely with the strong correlation,
determined from galaxies with inactive black holes, between supermassive
black hole mass and spheroid velocity dispersion \citep{fer00,geb00}: 
QSOs occur in the sorts of galaxies 
known to have the most massive central black holes.

At the present epoch, only a tiny fraction of galaxies with
massive spheroids shows luminous QSO activity.  The very steep evolution of
QSO activity with redshift indicates that some additional
ingredient besides the mere presence of a supermassive black hole is
necessary to produce QSO activity, and that this ingredient was much
more common in the early history of the Universe.  It has often been
speculated that the mechanism underlying this evolution is the 
sudden inflow of
gas to the center brought about by strong interactions or mergers.  There
has long been a fair amount of circumstantial evidence to support this
idea \citep[see, \eg][and references therein]{sto99}, yet
such arguments are by no means conclusive.

The debate about the nature of QSO host galaxies presently centers on the 
question of how significant tidal interactions are for QSOs generally:  
Do {\it most} QSOs at the current epoch begin their lives as mergers, 
or do most QSOs simply occur in old ellipticals to which nothing very 
interesting has happened recently?

We are conducting a coordinated study with Keck spectroscopy and Hubble
Space Telescope ($HST$) imaging of classical QSO host galaxies to investigate 
whether such hosts are truly quiescent ellipticals with ancient stellar
populations, or whether they are the results of mergers in the more recent 
past and have assumed elliptical morphologies only as a result of violent
relaxation due to the mergers.

Elliptical hosts formed through mergers would be expected to
show fine structure indicative of past tidal interactions, such as
shells and ripples. 
Studies of nearby merger remnants \citep[\eg][]{sch92,sch90}
indicate that such structure can in general be detected
even a few Gyr after the last major merger event.

To look for any potential fine structure, 
we recently obtained very deep $HST$ Advanced Camera for Surveys (ACS) images 
in a pilot study of five classical QSO host galaxies.  In this paper,
we present results for the first object, MC2\,1635+119.  The remaining four 
objects will be discussed in a subsequent paper 
(Bennert et al., in preparation).

The host galaxy of MC2\,1635+119 ($z = 0.146$;
1\arcsec $\simeq$ 2520 pc for $\Omega_{\Lambda}$ = 0.7, 
$\Omega_{\rm matter}$ = 0.3, and
 H$_0$ = 71 km\,s$^{-1}$\, Mpc$^{-1}$) was first described
by \citet{hut88} as having ``slightly elliptical amorphous structure'' 
with a luminosity profile that does not follow a simple exponential 
or $r^{1/4}$ law. 
Several companions are seen in the optical images \citep{hut88, mal84} as 
well as in the IR \citep{dun93}, without
any clear signs of interaction \citep{hut88}. \citet{mcl99}
compare fits to the host galaxy using an exponential disk and a de 
Vaucouleurs spheroid model, and conclude that the host resembles more 
closely an elliptical.
Regarding the stellar contents, 
\citet{nol01} estimate an age of 12 Gyr for the dominant stellar population
in the host galaxy from off-nuclear spectra.  

Thus, previous studies seem to indicate that the galaxy hosting 
MC2\,1635+119 is an elliptical with an old stellar population.   We now 
present new $HST$ and Keck observations that are in stark contrast with 
any such conclusions.

\section{Observations and Data Reduction}\label{observations}
Spectroscopic observations and their analysis are described in detail elsewhere
(Canalizo \& Stockton, in preparation).   Briefly, we obtained a
spectrum of the host galaxy of MC2\,1635+119 with
a total exposure time of 1.5 hours using the  Low-Resolution 
Imaging Spectrometer (LRIS; \citealt{oke95}) on the 
Keck I telescope on 2002 March 4.  
We used the 400 groove mm$^{-1}$ grism blazed at 3400\,\AA\ for the blue 
side (LRIS-B), and the 300 groove mm$^{-1}$ grating blazed at 5000\,\AA\ 
for the red side (LRIS-R), yielding dispersions of 1.09\,\AA\ pixel$^{-1}$ 
and 2.55\,\AA\ pixel$^{-1}$ respectively.
The slit was 1\arcsec\ wide, projecting to $\sim$7 pixels on the UV- and 
blue-optimized CCD of LRIS-B and $\sim$5 pixels on the Tektronix 
2048$\times$2048 CCD of LRIS-R.  The slit position angle (PA) was 57$\degr$,
placed roughly along the semi-major axis of the host galaxy and going 
through the QSO nucleus.  The object was observed near transit, so that the
effects of differential atmospheric refraction were minimized.  

The host galaxy spectrum was reduced using standard procedures.  A scaled
version of the QSO spectrum was subtracted from that of the host galaxy; the
spectrum was scaled by measuring the amount of flux in broad lines in the
spectrum of the host.  The final spectrum corresponds roughly
to a region 2$-$5$\arcsec$ from the nucleus on either side of the QSO and
has a signal-to-noise ratio $\sim$20.  
The spectrum was then modeled by performing least-squares fits to the 
data using preliminary \citet{cb07} and \citet{mar05}
population synthesis models as described in \S~\ref{pops}.  Both the models
and the observed spectrum were rebinned to the same spectral resolution.

Imaging observations were obtained using 
ACS/WFC onboard the $HST$
with the broad V-band F606W filter ($\Delta \lambda$ = 2342\AA; 1 pixel
corresponds to 0.05\arcsec).  We obtained
five sets of dithered images, each with four subsets of 550-586 s exposures,
yielding a total integration time of 11432~s.

\begin{figure}[bht]
\onecolumn
\epsscale{1}
\plotone{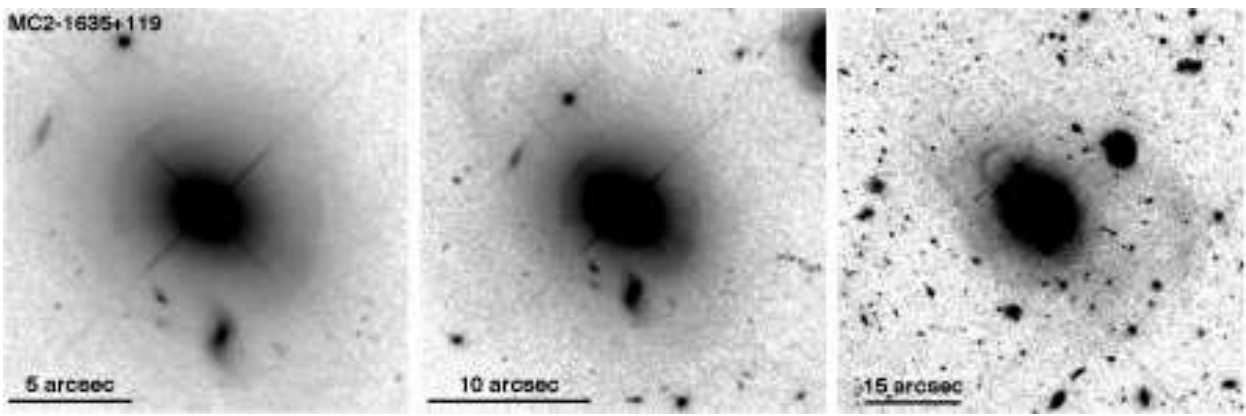}
\caption{ACS/WFC image of MC2\,1635+119, shown at different scales.   Fine 
structure consisting of shells,
arcs, and other debris is clearly seen at small and large scales. The images 
have been Gaussian smoothed with a sigma of either 0.5 pixels (left
and central panels) or 2 pixels (right panel). In this 
and the following figures, north is up and east is to the left.}\label{final}
\twocolumn
\end{figure}

We re-calibrated the data manually, starting from the pipeline flat-fielded 
individual exposures to improve the bias subtraction,
i.e., to correct the offset (of a few DNs) between the adjacent quadrants
that is still present in the final product of CALACS \citep{pav04}.
We then used MultiDrizzle \citep{koe02} to combine the individual images,
using the default values, bits=8578, as well as a deltashift-file
containing the offsets between the images as determined from stars
within the field-of-view (FOV).   The final distortion-corrected image is
shown in Fig.~\ref{final}, where the host galaxy shows clear shell structure.


\section{Image Processing}

\begin{figure}
\epsscale{0.85}
\plotone{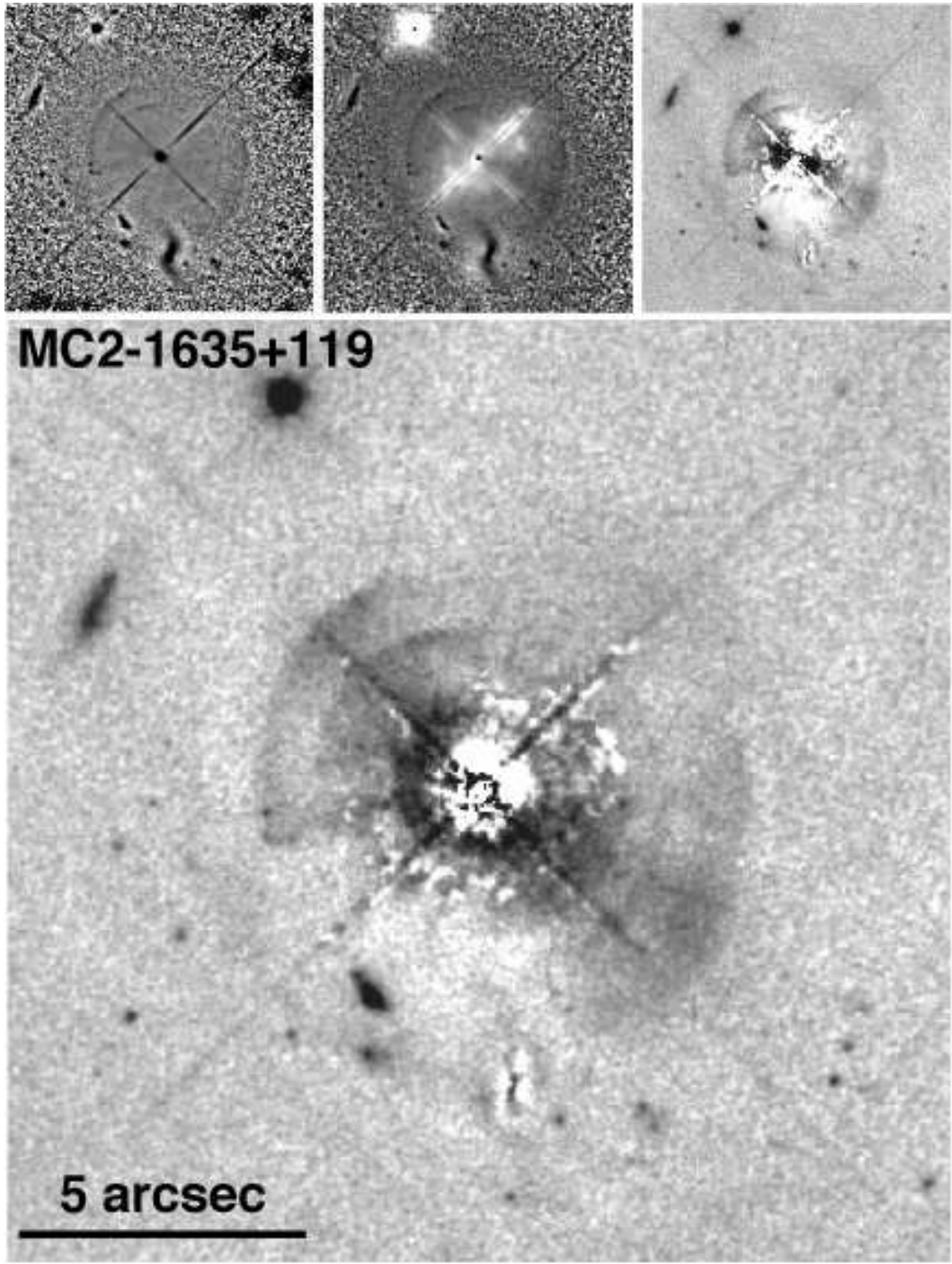}
\caption{Different methods used to detect fine structure in MC2\,1635+119,
as described in the text.   {\em Top left:} an unsharp-masked image, 
$f_{\rm unsharp}$.  {\em Top middle:} a structure map, $f_{\rm structure}$.
{\em Top right:} a residual image using GALFIT, where the model used for the
host galaxy consists of a de Vaucouleurs profile only.
{\em Bottom:} a residual image using GALFIT, where the model used for the
host galaxy consists of a de Vaucouleurs and a S{\'e}rsic profile of 
index n$\sim$1.
}\label{methods}
\end{figure}

To enhance and analyze any fine structure that might be present,
we applied various methods such as
unsharp masking, creating a so-called structure map
\citep{pog02}, as well as subtracting a central point spread function 
(PSF) for the QSO and a host galaxy model
making use of GALFIT \citep{pen02}.
All different approaches confirm the existence
of distinctive shells in the host galaxy (Fig.~\ref{methods}),
and we discuss each of them in turn.

To create an unsharp-masked image, we divided the final image $f$
by the $f$ convolved with a Gaussian function 
of $\sigma$ = 5 pixel ($G$): 
\begin{eqnarray*}
f_{\rm unsharp} = \frac{f}{f \otimes G}
\end{eqnarray*}

The structure map was derived by dividing $f$ 
by the PSF-smoothed image ($f$ $\otimes$ $P$)
and then convolving this ratio with the transpose of the PSF ($P^t$):
\begin{eqnarray*}
f_{\rm structure} =  \left[\frac{f}{f \otimes P} \right] \otimes P^t
\end{eqnarray*}
This process enhances unresolved or slightly resolved structures
on the scale of the PSF by removing the smooth light distribution
on larger scales \citep{pog02}.


\onecolumn
\twocolumn

A PSF image is needed for both the structure map and for modeling with
GALFIT.  Therefore, we created an artificial PSF from
TinyTim (Version 6.3) at the same position as our object as well
as a ``real'' PSF using a star on an ACS/WFC F606W image.
This image was obtained by searching the $HST$
archive for a suitable star 
at roughly the same chip position as the QSO and 
with a high signal-to-noise ratio (S/N). We found a star at
$<$ 30 pixels away from the position corresponding to the QSO 
with a S/N of 20000
that was observed on 20 dithered images
with a total exposure time of $\sim$8100s
(GO-9433, datasets j6mf19* and j6mf21*).
We processed these images
in the same manner as described
above for our data.

In order to minimize introducing additional noise into the PSF
subtraction and convolution operations, we first eliminated a few 
faint objects surrounding the PSF.  Then, 
depending on the data values compared to the standard deviation $s$
of the surrounding sky, we modified the PSF image as follows:
(1) for data values $>7s$, we retained the unmodified PSF; (2) for data
values between $3s$ and $7s$, we smoothed the image with a Gaussian
kernel with $\sigma = 0.5$ pixel; (3) for smaller data values, we
smoothed with a Gaussian kernel with $\sigma = 2.0$ pixel; finally, (4)
for data values that were $<s$ after this last operation, we replaced
the value with 0.

To probe the quality of the two different PSFs,
we subtracted them from both saturated and unsaturated
stars within our FOV using GALFIT: the
real PSF star gave significantly better results than
the TinyTim PSF.  From this exercise, we also determined
the central region with a radius of $\sim$1\farcs7 is 
strongly affected by the PSF 
subtraction because the QSO nucleus was saturated; any structure
seen within this region is likely an artifact.

The best enhancement of the shell structure was obtained
using GALFIT (Fig.~\ref{methods}), a 2-dimensional galaxy fitting program
capable of fitting simultaneously
one or more objects in an image with
different model light distributions \citep[such as \citet{ser68}, 
\citet{deV48},
exponential, etc.;][]{pen02}. 
Briefly, our adopted procedure was as follows:
First, we created a mask to exclude the saturated pixels in the center,
the diffraction spikes, any surrounding bright
objects, and the shells themselves, in order to fit only
the smooth underlying host galaxy light distribution.
Then, a (``real'') PSF as well as several
S{\'e}rsic functions were fitted.
In GALFIT, the S{\'e}rsic power law is defined as
\begin{eqnarray*}
\Sigma (r) = \Sigma_e \exp \left[- \kappa \left(\left(\frac{r}{r_e}\right)^{1/n}-1\right)\right] \hspace{0.2cm} 
\end{eqnarray*}
where $\Sigma_e$ is the pixel surface brightness at the effective radius $r_e$ \citep{pen02},
and $n$ is the S{\'e}rsic index ($n=4$ de Vaucouleurs, $n=1$ exponential
profile).
In addition, we fitted
the bright neighbor to the south of the QSO with a S{\'e}rsic function.
In all steps, the background sky was fitted simultaneously.
This least-squares fit was then subtracted
from the original image to gain the residual image, enhancing all structure
that lies on top of the smooth host galaxy light distribution.

When we used a single component for the host galaxy, the best fit was 
achieved with 
a S{\'e}rsic function of index n = 8.8.   This fit was marginally better 
(only a few percent in $\chi^2$) than the fit achieved using a de 
Vaucouleurs profile.  On the other hand, the fit resulting from an 
exponential profile was much worse (roughly 50\% in $\chi^2$).  This 
finding is in agreement with
the results by \citet{mcl99}, who determined that the host
galaxy of MC2\,1635+119 is better fit
by a de Vaucouleurs than an exponential profile.

The fit improved substantially, however, when two components were included
instead of one.  Using two S{\'e}rsic functions, the best result was 
achieved when one had an index n=4, 
which corresponds to a de Vaucouleurs profile, and the other an index n=0.91, 
which corresponds nearly to an exponential disk; this fit is shown in 
Fig.~\ref{methods} and listed in Table~\ref{results}. If, instead, the 
index of one of the S{\'e}rsic 
components was fixed to n=1 (exponential), the best fit was achieved when 
the other component was close to a de Vaucouleurs profile (with index n=4.6;
Table~\ref{results}).
Therefore, we conclude that the host galaxy is well modeled by a de 
Vaucouleurs spheroid plus an exponential disk that makes up roughly one 
fourth of the light in the surface brightness profile, as detailed
in Table~\ref{results}.   In that table, we also list results for the fit 
using a de Vaucouleurs profile only in order to compare our results 
with those of \citet{dun03}, and we find that our results are 
very 
similar to theirs.
However, as Fig.~\ref{methods} shows, the resulting 
model-subtracted image using only a de Vaucouleurs profile has residuals
that are significantly larger than those obtained when we use a 
two-component model.

\begin{deluxetable}{lccccccc}
\tabletypesize{\footnotesize}
\tablecolumns{6}
\tablewidth{0pc}
\tablecaption{Results of modeling the QSO host galaxy using GALFIT}
\tablehead{
\colhead {fit type} &
\colhead{function} & \colhead{$\Delta$($\alpha$,$\delta$) (arcsec)} & \colhead{$m_{\rm F606W}$ (mag)} & \colhead{$r_e$ (kpc)} & \colhead{S{\'e}rsic index} & \colhead{$b/a$} &
\colhead{PA (deg)}\\
\colhead{(1)} & \colhead{(2)} & \colhead{(3)}  & \colhead{(4)} & \colhead{(5)}
& \colhead{(6)} & \colhead{(7)} & \colhead{(8)}}
\startdata
de Vauc+S{\'e}rsic & S1 & ($-$0.03,0.04) & 17.46 & 2.74 & 4 (fixed) & 0.74 & 57.3\\
 & S2 & (0.72,0.26) & 18.80 & 15.89 & 0.91 (free) & 0.79 & 28.8\\
\hline
S{\'e}rsic+Exp & S1 &($-$0.03,0.04) & 17.39 & 2.75 & 4.6 (free) & 0.75 & 57.3\\
               & S2 & (0.85,0.29)  & 18.83 & 16.5 & 1 (fixed) & 0.79 & 26.9\\
\hline
de Vauc+Exp & S1 & ($-$0.03,0.04) & 17.45 & 2.68 & 4 (fixed) & 0.74 & 57.4\\
& S2 & (0.72,0.26) & 18.71 & 16.03 & 1 (fixed) & 0.8 & 29.3\\
\hline
de Vauc only & S1 & ($-$0.03,0.04) & 17.26 & 5.74 & 4 (fixed) & 0.75 & 52.3\\
\hline
\citet{dun03} & S1 & \nodata & \nodata & 5.73 & 4 (fixed) & 0.69 & 56
\enddata
\tablecomments{
Column (1) lists the GALFIT model, (2) the individual components used 
(S = S{\'e}rsic), (3) the offsets with respect to the PSF,
(4) the integrated apparent magnitude in the F606W filter,
(5) the effective radius, (6) the S{\'e}rsic index, (7) the
axis ratio, and (8) the position angle (east of north).
Results from \citet{dun03} are listed for comparison.
Note that the PA given here for the \citet{dun03} results was derived by
adding the orientation of the spacecraft to the PA given in their Table 3,
which was apparently not corrected for this orientation.
}
\label{results}
\end{deluxetable}

\section{Shell Structure and Luminosity}\label{shells}

\begin{figure}[bht]
\onecolumn
\epsscale{1}
\plotone{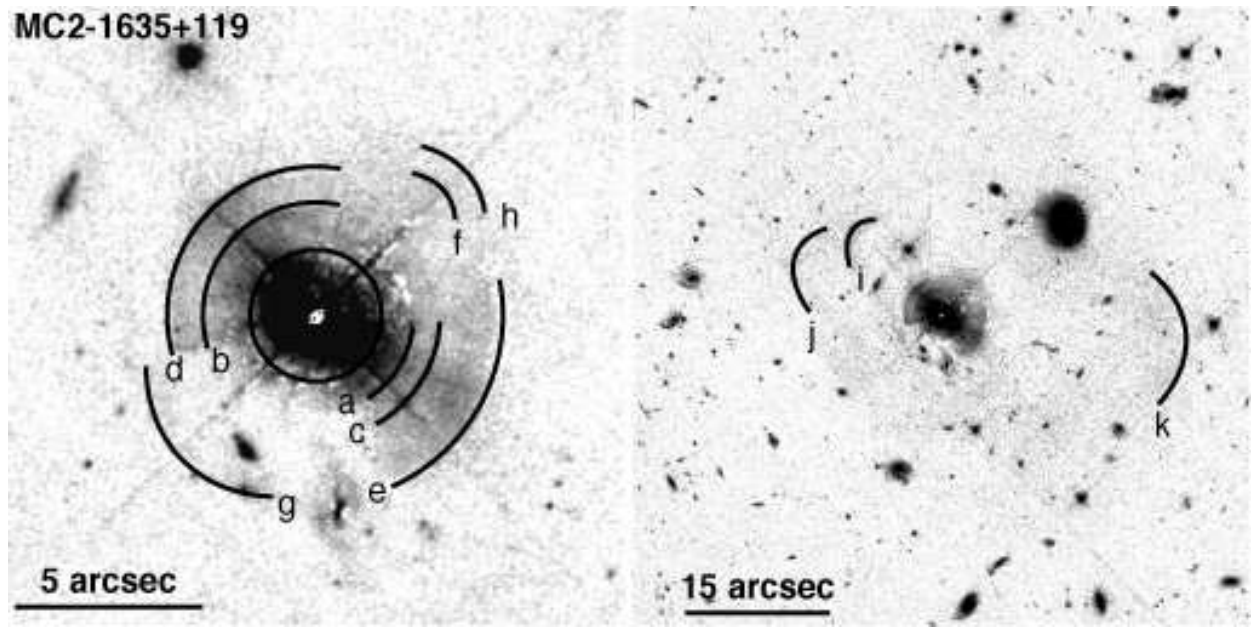}
\caption{Model-subtracted images of MC2\,1635+119, where the most prominent
fine-structure features are labeled.
}\label{shell_labels}
\twocolumn
\end{figure}

In Fig.~\ref{shell_labels} we show a residual image of MC2 1639+119 indicating
the position of the different tidal features that we identify.  The central
circle with a radius of 1\farcs7, corresponds to the area most affected by the
saturated PSF; any features within this area may be artifacts of the 
PSF subtraction.  Unfortunately, this prevents us from reliably 
detecting any shells or other structure that may be present in that region.

The arcs labeled $a$ through $e$ in Fig.~\ref{shell_labels} are all segments
of circles centered on the galaxy, emphasizing the regularity of the
interleaved shells. The projected radii of these shells are roughly 6.6, 
7.6, 8.3, 10.0, and 12.5 kpc, respectively.  This set of bright shells 
is closely aligned with the semi-major axis of the host galaxy, 
at PA $\sim$54$\degr$.  The shell system shows roughly a 
biconical structure, although the edges of this putative bicone do not 
intersect at the center of the 
host galaxy. Shell $e$ shows a discontinuity west of the QSO that may be 
due to obscuration by dust, or the shell may be made of two or more 
components.

A set of lower surface-brightness shells or ripples ($f$, $g$ and $h$) with
seemingly different (greater) ellipticities are seen roughly perpendicular to 
the first set, both north-west and south-east of the nucleus.

Further
out to the north-east, there is an arc-like feature ($i$) extending out to
a projected distance of $\sim$32 kpc.    Other faint tails or wisps are seen 
in that same 
region ($j$).   Finally, a much larger, faint and diffuse feature resembling 
either a shell or some tidal tail ($k$) is visible $\sim$65 kpc west of the 
nucleus.  While this feature is very faint, we are
confident it is real, particularly as this feature is also visible in a WFPC2 
archival image (GO-6776) when the image is median filtered and 
Gaussian smoothed.
 
To estimate the luminosity within the shells compared
to the total luminosity of the galaxy, we
created a mask that includes all the light
within an annulus of inner radius of 1\farcs7
and outer radius of 6\farcs8, but that at the same time excludes
\vspace{4.5in}
the diffraction spikes as well as several additional light sources
from apparent companions.
Note that the outer radius was chosen to be 3 $\times$ $r_{\rm eff}$
with $r_{\rm eff}$ determined from a single de Vaucouleurs fit 
(see Table~\ref{results}).
This mask (good=1, bad=0) was multiplied by the image and the total 
counts in the product were summed. This was done for both 
the GALFIT residual image
($f_{\rm shells}$) obtained subtracting the GALFIT model
of a de Vaucouleurs + exponential profile,
and the GALFIT model itself ($f_{\rm galaxy}$).
Finally, we computed the ratio $f_{\rm shells}$/$f_{\rm galaxy}$.
This yields the fractional luminosity of the shells between
1\farcs7 and 6\farcs8 radius as $\sim$ 6\% of the host galaxy
light (within the same annulus).  This estimated percentage may be smaller 
or larger than the true percentage
depending on whether there are any shells within the radius affected by the PSF subtraction or not.

Note that the percentage given refers to the total flux within the shells 
out to 13 kpc.
However, the local contrast between the shells
and the galaxy (as estimated by
dividing the residual image by the GALFIT model)
varies between 5 and 20\% and reaches 50\%
in Shell $e$.

\section{Time Constraints from Tidal Structure}\label{timescales}

As described above, the host galaxy of MC2 1635+119 reveals spectacular 
structure of regular and aligned shells on projected radii of 5-13 
kpc. Similar shells are observed in some local giant ellipticals
\citep[\eg][]{mal83,sch80,sik07} and are interpreted as remnants of a 
merger event.   It has been shown that the mergers that produce shell-like
structure can be either minor \citep[][hereafter Q84]{q84} or major 
\citep[\eg][]{her92}.  In this 
section, we discuss both scenarios in the context of the morphology and 
physical size of the shells and structure we detect, with the aim of placing 
constraints on the age of the tidal interaction that formed them.

\subsection{Minor Merger}\label{minor}

We first consider the case of a minor merger since it allows for the simplest
physical interpretation of the data.  In this scenario, the system of regular
concentric shells, confined within a finite range in azimuth, can result from 
the merger of a smaller galaxy (either spiral or elliptical) with a large 
elliptical along a nearly radial orbit
\citep[Q84;][hereafter HQ88 and HQ89]{dup86,hq88,hq89}.

The shell formation mechanism works as follows: during the merger, 
stars from the smaller galaxy are captured by the massive galaxy 
and start to oscillate in its potential well. Since stars spend most 
of the time near the apocenters of their orbits (where their radial 
velocities go 
to zero), a relative enhancement of the stellar density (a shell) forms 
there. The first shell is formed by captured stars that were initially in 
orbits with the 
smallest oscillation period, \ie\ those with the smallest apocenter distance.

As time goes on, the shortest-period stars move away from apocenter, 
while stars with slightly longer periods reach their apocenter at a 
slightly larger galactocentric distance. Due to a 
continuous range of oscillation periods, the first shell appears to 
propagate radially outward while its stellar content progressively changes: 
it is thus a radially propagating stellar density wave. A new traveling 
shell appears every time the shortest-period stars complete another 
oscillation period. After several oscillations, the massive elliptical 
galaxy reveals a system of shells where the outermost shell is the oldest,
since this is the shell that formed first.
This scenario gives a simple relation between the radius of this shell 
and the time of its formation.

We have constructed a simple N-body model that reproduces, at least 
qualitatively, the brightest shells observed in MC2\,1635+119. The N-body 
model uses the same technique as that used by Q84 and HQ88. 
In this model, the secondary (smaller galaxy) moves on a radial orbit and 
is assumed to be
disrupted instantaneously by the tidal forces of the primary 
(massive elliptical) after the first
passage through the center of the primary. 
This corresponds to 
abruptly lowering the secondary's mass to zero, after which the test particles
move in the potential of the primary alone.  
Thus, dynamical friction is assumed to be unimportant, and the model should 
only be considered as a zero-order description of the collision.

We assumed a radial orbit with an initial separation between galaxies
arbitrarily chosen to be 90 kpc (5-18 times the scale-length of the primary).
The initial infall velocity of the secondary was set equal to the 
escape velocity for the potential of the primary.

We simulated the merger using (1) a de
Vaucouleurs profile, and (2) a Plummer sphere (corresponding to a 
Moffat's n=2 surface brightness profile). 
Since the goal of these simulations 
was to provide only a first order estimate of the merger 
timescale, we did not attempt to use more realistic composite 
density profiles of luminous and dark matter.
We used effective radii
ranging from 5 to 20 kpc; this range spans values for $r_{\rm eff}$ 
found by \citet{dun03}, 
\citet{tay96}, and our own work (all corrected to the cosmology used in this 
paper).
The mass of the giant elliptical was taken to be 
3.2$\times$10$^{11}$~M$_{\odot}$
\citep{dun03}, although we allowed for a range of masses up to 
3.2$\times$10$^{12}$~M$_{\odot}$ in order to account for a dark matter halo.
The secondary-to-primary mass ratio and scale-length ratio were both fixed 
to 0.1; we note that while the precise choice of these two ratios is 
arbitrary, they affect mainly the contrast of the shells, and not the 
timescales, as long as the primary dominates the potential.   

\begin{figure}
\epsscale{0.9}
\plotone{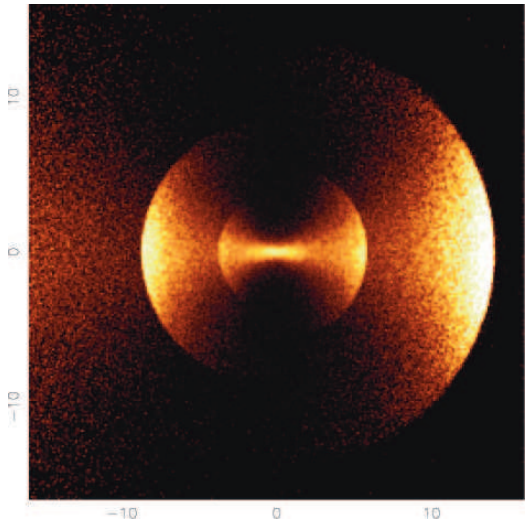}
\caption{Shell structure in a restricted N-body simulation of a minor merger 
of two ellipticals (gE+dE).  The masses of the galaxies 
are, respectively, $3.2\times10^{11}$~M$_{\odot}$ and 
$3.2\times10^{10}$~M$_{\odot}$, and their 
effective radii 5 and 0.5 kpc.  Both galaxies are modeled as Plummer spheres.
The smaller galaxy came from the right on a radial orbit. 
The box is 16$\times$16 kpc.  Only the particles belonging to the smaller 
galaxy are shown, to allow for comparison with images where a model of the
host galaxy has been subtracted.
}\label{numerical}
\end{figure}

Figure \ref{numerical} shows our results for simulations
using a Plummer surface brightness profile. 
The de Vaucouleurs model, which leads to lower contrast and more spherical 
shells, will be discussed in
more detail in a subsequent paper (Jungwiert et al., in preparation).
Table~\ref{modeltable} lists the timescales for two outermost shells 
(see below) to reach their observed radii in models with the 
range of parameters for the primary given above. We measure this timescale
from the moment when the centers of mass of the two galaxies pass by each other
(hereafter ``merger timescale'').
We do not attempt to use the sizes or separations of inner shells to constrain 
the timescale since inner shells are more sensitive to the exact shape of 
the central density profile of the primary and are also more likely to be 
influenced by dynamical friction, which is not implemented in our model.

\begin{deluxetable}{cccc}
\tablecolumns{4}
\tablewidth{0pc}
\tablecaption{Shell Formation Timescales from Numerical Simulations}
\tablehead{
\colhead {$R_{shell}$} & {$M_{primary}$} & {$T_{deVauc}$} & {$T_{Plummer}$} \\
\colhead{(kpc)} & {(M$_{\odot}$)} & \colhead{(Myr)} & \colhead{(Myr)}
}
\startdata
12.5 & 3.2$\times$10$^{11}$ & 100 $-$ 245 &    145 $-$ 400 \\
 & 3.2$\times$10$^{12}$ & \phn30 $-$  \phn60 &   \phn45 $-$ 135\\
\hline
65 & 3.2$\times$10$^{11}$ & \phn360 $-$ 1720 &    1380 $-$ 1620 \\
 & 3.2$\times$10$^{12}$ & \phn100 $-$  \phn400 &   \phn440 $-$ \phn510\\
\enddata
\tablecomments{The time range given for each model corresponds to a range of
effective radii for the giant elliptical of 5 to 20 kpc.  The time is measured
from the moment when the centers of mass of the two galaxies pass through
each other.}
\label{modeltable}
\end{deluxetable}

Table 2 shows that, allowing for an uncertainty in the type of profile, for 
a rather large uncertainty in the effective radius, and for a considerable
amount of dark matter, the time for Shell~$e$ to reach its present distance
of 12.5 kpc appears to be confined to a range of $\sim$30$-$400 Myr after
the centers of the two galaxies passed through each other.  

These ages are calculated assuming that Shell~$e$ is the outermost shell.  
However, we might consider the possibility that the tidal feature $k$ may be
a much older, fainter shell formed during the same encounter.   This ``shell'',
at a projected distance of $\sim$65 kpc from the center of the host galaxy, 
would then give a merger timescale ranging from 100 Myr to 1.7 Gyr 
(see Table~\ref{modeltable}), given the assumptions considered above. 

We emphasize that our simulations model the simplest plausible case, and 
at this point we cannot exclude more complicated scenarios.  In a subsequent 
paper (Jungwiert et al., in preparation) we will consider N-body simulations 
of this galaxy and of shell galaxies in general in more detail, focusing 
on different gravitational potentials, various mass ratios 
of colliding galaxies, dynamical friction, 
tidal stripping and the fate of gas.

\subsection{Major Merger}

While the numerical simulations described above can reproduce the
morphology of the brightest shells in MC2\,1635+119, they do not rule out the 
possibility that the shells might have been created by a major merger.  
Further, the 
model-subtracted images (Figs.~\ref{methods} and \ref{shell_labels}) show 
features ($f$, $g$, $h$) that are off-axis from the direction of the 
encounter implied by the inner shells. 
Additional tidal debris at different
position angles is seen on much larger scales (features $i$, $j$, $k$).  
It is difficult to explain how all this structure might have formed 
as a result of a minor merger,
provided a single interaction is responsible for all the features.

The fact that the inner shells appear to be closely aligned with the major
axis of the host would also argue against a minor merger 
\citep[see][and references therein]{her92}.  Using numerical simulations, 
\citet{her92} show that mergers between two disk galaxies of similar mass 
can form shells, loops, and ripples.   In particular, their simulations are 
compared to NGC\,3923, one of the best examples of a nearby 
elliptical galaxy with shells \citep{mal83}.  
The system of shells of NGC\,3923 ($z=0.005801$) 
extends from distances close to the center ($<$2 kpc)
out to $\simeq$100 kpc \citep{pri88}. The shells are distributed roughly in 
an hour-glass shape with an opening angle of $\simeq$ 60$\degr$. While most 
of the shells appear aligned with the major axis of the galaxy, the outermost 
shell does not, a feature that is nicely reproduced by the 
simulations by Hernquist and Spergel.   These characteristics are 
similar to those
observed in MC2\,1635+119, although it should be noted that the structure
of the inner shells in MC2\,1635+119 is significantly more regular 
(non-intersecting and aligned) than
that of the NGC\,3923 shells or of the numerical simulations by 
\citet{her92}.   However, 
the comparison does point out that a major merger
could also have formed the shells seen in MC2\,1635+119.

The amount of light observed in the shells may yield further clues to the 
nature of the merger.   As mentioned in \S\ref{shells}, the system
of five bright shells comprises $\sim$6\% of the total luminosity of the
galaxy.   However, the shells contain only a fraction of the total number of
stars that were originally part of the merging galaxy, \ie\ those whose 
orbital velocities are near zero.   Our numerical simulations and those of
\citet{her92} indicate that the stars in shells make up only one fourth or 
less of the total mass of the companion.   Therefore, assuming that the 
mass-to-light
ratio is similar in both galaxies, the intruder may make up about 24\% of
the total mass.   If we add to that the mass implied by the more extended
``shell'', the fraction may be closer to 30\%.  Thus, by this argument alone, 
the mass ratio of the original galaxies may have been close to 7:3 which 
may be considered a borderline major merger.

Our simple N-body model produces shells up to a mass 
ratio of 3:1 for the parent galaxies.  We did not investigate smaller mass
ratios due to the increased complexity of such encounters.  If we assume
that the ``shell'' at 65 kpc (Shell $k$) formed through a similar mechanism 
as that outlined in \S\ref{minor}, then the range of timescales of 100 Myr 
to 1.7 Gyr would still hold for a major merger.  If, however, this feature 
was formed through the spatial wrapping of, \eg\ a tidal tail, then estimating
a timescale becomes more complex since timescales become more 
heavily dependent on
initial conditions.  As a reference, we note that simulations of the 
major merger in ``The Mice'' by \citet{bar04} produce a merger remnant 
somewhat similar to MC2\,1635+119 at a time close to 1 Gyr from the beginning
of the merger event.

\section{Stellar Populations}\label{pops}

\begin{figure}[bh]
\epsscale{0.9}
\plotone{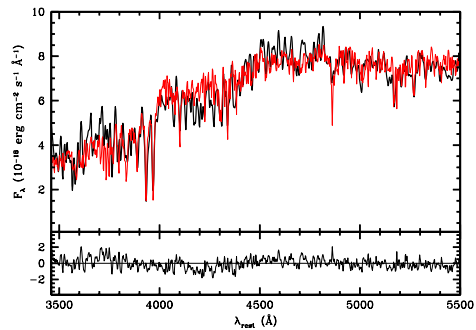}
\caption{Keck LRIS spectrum of the host galaxy of MC2\,1635+119 in rest frame.
The black trace is the observed spectrum.  The red trace is the best fit 
\citet{cb07}
model to the data. The model consists of 52\% (by mass) of a 1.4 Gyr old 
population and 48\% of a 12 Gyr population.  In the bottom panel we show the 
residuals obtained by subtracting the model from the observed spectrum.
}\label{spectrum}
\end{figure}

Figure~\ref{spectrum} shows the Keck LRIS spectrum of the host galaxy of 
MC2\,1635+119 in rest frame, representing its integrated light from 2\arcsec\
to 5\arcsec\ radius along the slit on either side of the nucleus 
(see \S\ref{observations}). Since the slit was placed roughly in the 
direction of the major axis of the host galaxy, the spectrum includes 
the brightest shells in the host (Fig.~\ref{shell_labels}, $a$ through $e$). 

The stellar component has a redshift 
$z_{\rm abs}=0.1474$ (measured from absorption lines), equal to the redshift
we measure from narrow 
emission lines, but slightly higher than that of the broad emission lines
($z\sim0.146$).   

In order to model the spectrum, we used population synthesis models by
\citet{mar05} and the preliminary models by \citet{cb07}.
We chose these two sets of models because they provide the best
match to our spectral resolution and they both include contributions
from thermally pulsating asymptotic giant branch stars, which are known to be
particularly important in intermediate-age ($\sim$1 Gyr) stellar populations
\citep{mar05}.
Our original approach to analyzing the spectrum was
to assume a dominant old stellar population ($\sim$12 Gyr) representing  
the population of the giant elliptical galaxy, with a smaller fraction of more 
recent star formation possibly triggered by the merger that formed
the shells.   Models that include a very small fraction 
($<$ 0.3\%) of a young ($<$ 50 Myr) starburst and a dominant ancient
population can produce a rough fit to the continuum, but the fit to
individual features such as Ca\,II~H\&K and the CN band is rather poor.
We tested spectral fits using different metallicities ranging from 0.02 to 
2 solar, and found that solar metallicity models consistently yielded the 
lowest $\chi^2$.   The choice of initial mass function 
\citep{cha03,kro01,sal55} made little or no difference.

However, the best fit to the observed spectrum, including both the continuum 
and stellar features, was achieved by adding a large contribution
from an intermediate-age starburst population to the 12 Gyr model.  
A better fit was
achieved with \citet{cb07} than with \citet{mar05} models, but both sets of 
models
yielded similar results.   In the case of \citet{cb07} models, the best fit 
(shown in Fig.~\ref{spectrum}) corresponds
to an intermediate-age population of 1.4 Gyr contributing 52\% of the total 
mass along the line of sight.   The best fit using \citet{mar05} models is 
for an intermediate-age population of 1.0 Gyr contributing 45\% of the total 
mass along the line of sight.   The real difference between the two models
may be 
even smaller, considering that the Maraston models provide a coarser age grid 
(with steps in age at 1.0 and 1.5 Gyr)
than the Charlot \& Bruzual models and the fact that the $\chi^2$ for 
the latter
shows a shallow minimum from $\sim$1.2 to 1.9 Gyr (although the mass
contribution from the starburst increases steeply with age). In both cases, 
$\chi^2$ increases rapidly beyond 2.0 Gyr.   

The determination of these intermediate-age components is robust with 
respect to the 
choice of the model for the older population: the same intermediate-age 
populations are obtained when the older population is varied from 6 to 14 Gyr.
If we use models of metallicities lower than solar, a single population can be 
used to fit the data, although the overall fit is significantly worse.   
In this case, the 
oldest population that yields a reasonable fit is less than 3 Gyr old.   
Single populations older than 4 Gyr yield poor fits regardless of 
their metallicity or initial mass function.   Although it is 
possible that the spectrum may be somewhat reddened by dust, it is unlikely
that the age of the starburst component would be significantly younger than
one Gyr, given the absorption features that we observe.
Finally, an inaccurate subtraction of the QSO contribution could affect the
shape of the continuum.  We tested the effects of this by fitting spectra that
were slightly over-subtracted and under-subtracted.  While the $\chi^2$ for 
these cases was somewhat larger, the age of the starburst for the best fit
remained the same.

Naturally, the number of possible combinations of populations to model the
spectrum of the host is large.   We have kept our analysis simple by
testing only a limited number of possibilities corresponding to physically
plausible scenarios.   Therefore, while we cannot exclude more complex
star-formation histories, we are fairly certain that:
(1) The dominant component of the stellar population in the host of 
MC2\,1635+119 is {\it not} ancient, and
(2) A small percentage by mass of recent (less than a few hundred Myr) star 
formation superposed on an old ($>6$ Gyr) population can be ruled out, 
regardless of the age of the dominant population.
Instead, the spectrum of the host of MC2\,1635+119 is dominated (at least in 
flux) by an intermediate age population of 1--2 Gyr.

\section{Discussion}

In agreement with previous observations, we have found that the surface 
brightness profile of the galaxy hosting MC2\,1635+119 is closer to a 
de Vaucouleurs than an exponential profile.  However, our new ACS image reveals
that a fainter exponential profile is also present, comprising up to one
forth of the total luminosity.   Moreover, our observations have uncovered a 
spectacular system of shells and other faint structure in the host galaxy
at small and large scales, showing that the host is far from being undisturbed.
We have also found that the stellar populations in the host galaxy seem to
have a substantial contribution ($\sim$50\% by mass) of an 
intermediate-age stellar population from a 1--2 Gyr old starburst.

While the large contribution of an intermediate-age population 
to the spectrum of the host galaxy of MC2\,1635+119 is intriguing, it is by
no means unusual.  Recent studies of AGN host galaxies \citep[e.g.,][]
{jan04,san04,kau03,can06} indicate that galaxies hosting the most luminous 
AGN are often dominated by bulges whose colors
are significantly bluer than those of inactive elliptical galaxies and
are consistent with the presence of intermediate-age starbursts.   Based on
positions of the hosts in the $D_{n}$(4000)/H$\delta_A$ plane, 
\citet{kau03} suggest that these AGN hosts have had significant bursts of 
star formation in the past 1--2~Gyr.  

Why do AGN host galaxies show these strong intermediate-age populations?
And, what is the physical connection, if any, between the putative $\sim$Gyr 
old starburst and the nuclear activity?   Understanding the nature of this 
relation is important because it could have implications for the triggering 
mechanisms and duty cycles of AGN.   Our study of MC2\,1635+119 provides some 
clues that may be applicable to a larger population.

We now know that the host galaxy of  MC2 1635+119 was unequivocally 
involved in a tidal encounter.  Our rough estimates discussed in 
\S~\ref{timescales} place the timescale for this encounter at less than
$\sim$1.7 Gyr, which could be compatible with the age of the major starburst.
However, the large uncertainty in our estimate does not rule out the 
possibility of a substantially more recent event.  We are also unable to 
discriminate between a major and a minor merger as the culprit for the 
shell structure that we observe.   Our results give us enough information, 
however, to speculate on a couple of likely scenarios.

First, consider the case where the inner shell structure was formed through
the accretion of a low-mass companion (one tenth or less of the mass of
the primary).   The overall morphology that we observe would have to be 
caused by more than one event, and the fact that there was a dramatic 
episode of star formation more than one Gyr ago would argue for a past (major?)
merger connected to the large-scale tidal debris.   In that case, it is
possible that the giant elliptical possessed a higher gas content as a result 
of the past merger event, and so the QSO activity was more readily triggered
(or rejuvenated) in it
by a minor merger than it would have been in a gas-poor elliptical would.   
This may well be the
case in Cygnus\,A, where an ongoing minor merger appears to be responsible for 
triggering the nuclear activity \citep{can03}. 

Consider now the alternative case where a major merger is responsible for 
both the 
starburst and all of the structure that we observe.  This merger event would
have occurred over one Gyr ago and would have likely (though not necessarily)
triggered a first episode of accretion onto the black hole(s).   Feedback 
from the QSO quenched any further star formation.  Assuming theoretical 
estimates for the duration of QSO activity are correct 
\citep[\eg\ 10$^{7} - 10^{8}$ yr;][]{yu02},
the QSO activity would have ceased as the merger continued 
its course and the morphology of the newly merged galaxies began to relax
into the shape of an elliptical.   Eventually, the extended tidal debris would
``rain'' back into the central regions of the galaxy, triggering a 
new episode of QSO activity.  
A time delay in the onset of QSO activity would be in agreement with 
predictions by hydrodynamical simulations of merging galaxies 
\citep[see \eg][]{bar98,spr05,hop07}.  These models frequently predict a 
second peak in star formation that also occurs much later in the merger. 
Since our spectroscopic observations exclude a radius of $\sim$5 kpc around 
the nucleus, we would not have detected any recent star formation that may
be present in the central regions of the host galaxy.

While these are interesting scenarios, they are, for the moment, no more 
than ``guided'' speculation.   More complete N-body models as well as high
angular-resolution spectroscopy to measure the kinematics of the stellar 
component are needed to get a better handle on the kind of encounter
that formed the observed structure.   However, we will also need to study 
larger samples to attempt to answer more complex questions, such as the precise
timing of the triggering of the QSO activity, which in turn should help
answer questions regarding duty cycles and feedback.

\acknowledgments
We thank S.~Charlot and G.~Bruzual for providing access to their
models prior to publication and the referee for helpful comments.  
Support for Program \# GO-10421 was provided by NASA through a grant from 
the Space Telescope Science Institute, which is operated by the Association 
of  Universities for Research in Astronomy, Incorporated, under NASA contract 
NAS5-26555.
Additional support was provided by the 
National Science Foundation, under grant number AST 0507450. 
B.J. acknowledges support by the Grant No.~LC06014 of the Czech Ministry 
of Education and by the Research Plan No.~AV0Z10030501 of the Academy 
of Sciences of the Czech Republic.
Some of the data presented herein were obtained at the W.M. Keck Observatory, 
which is operated as a scientific partnership among the California Institute 
of Technology, the University of California and the National Aeronautics and 
Space Administration. The Observatory was made possible by the generous 
financial support of the W.M. Keck Foundation.
The authors wish to recognize and acknowledge the very significant cultural role and reverence that the summit of Mauna Kea has always had within the indigenous Hawaiian community.  We are most fortunate to have the opportunity to conduct observations from this mountain.

{\it Facilities:} \facility{HST (ACS)}, \facility{Keck:I (LRIS)}

\end{document}